\documentclass[twocolumn,showpacs,%
  nofootinbib,aps,superscriptaddress,%
  eqsecnum,prd,notitlepage,showkeys,10pt]{revtex4-1}

\usepackage{amssymb}
\usepackage{amsmath}
\usepackage{graphicx}
\usepackage{dcolumn}
\usepackage{hyperref}
\usepackage{multirow}

\begin{document}

\title{Predicting Core Electron Binding Energies in Elements of the First Transition Series Using the $\Delta$-Self-Consistent-Field Method$^\dag$}
\author{J. Matthias Kahk}
\affiliation{Institute of Physics, University of Tartu, W. Ostwaldi 1, 50411 Tartu, Estonia.}
\author{Johannes Lischner}
\affiliation{Department of Materials, and the Thomas Young Centre for Theory and Simulation of Materials, Imperial College London, London SW7 2AZ, United Kingdom.}

\footnotetext{\dag~Electronic Supplementary Information (ESI) available: full numerical results, core-enhanced numerical basis sets used for the atoms with a core hole, an evaluation of the basis sets used in this work, relative to uncontracted gaussian basis sets, sample control.in files, and the relaxed structures of all molecules.}

\begin{abstract}

The $\Delta$-Self-Consistent-Field ($\Delta$SCF) method has been established as an accurate and computationally efficient approach for calculating absolute core electron binding energies for light elements up to chlorine, but relatively little is known about the performance of this method for heavier elements. In this work, we present $\Delta$SCF calculations of transition metal (TM) 2$p$ core electron binding energies for a series of 60 molecular compounds containing the first row transition metals Ti, V, Cr, Mn, Fe and Co. We find that the calculated TM 2$p_{3/2}$ binding energies are less accurate than the results for the lighter elements with a mean absolute error (MAE) of 0.73 eV compared to experimental gas phase photoelectron spectroscopy results. However, our results suggest that the error depends mostly on the element and is rather insensitive to the chemical environment. By applying an element-specific correction to the binding energies the MAE is reduced to 0.20 eV, similar to the accuracy obtained for the lighter elements.

\end{abstract}

\maketitle

\section{Introduction}

Theoretical calculations of core electron binding energies are important for aiding the interpretation of experimental core level X-ray photoelectron spectra. In recent years, it has been shown that the all-electron $\Delta$\nobreakdash-Self-Consistent-Field ($\Delta$SCF) method \cite{bagus_self-consistent-field_1965} can yield accurate absolute core electron binding energies in both free molecules and periodic solids, provided that computational details such as the choice of the exchange-correlation functional, the treatment of relativistic effects, basis set convergence, and size convergence in periodic calculations are carefully considered \cite{kahk_accurate_2019,kahk_core_2021,pueyo_bellafont_performance_2016-1,hait_highly_2020}. It has also been shown that eigenvalue self-consistent full-frequency GW calculations can yield accurate absolute core electron binding energies in free molecules \cite{golze_accurate_2020}.

However, thus far, almost all theoretical studies of core electron binding energies have focused on a narrow subset of elements in the periodic table. In particular, the prediction of 1$s$ core electron binding energies in the elements carbon to fluorine has attracted the most attention \cite{kahk_core_2021,ozaki_absolute_2017,kahk_accurate_2019,takahata_accurate_2010,aoki_accurate_2018,zhu_all-electron_2021,walter_offset-corrected_2016,kahk_core_2018,hait_highly_2020,regoutz_combined_2020,golze_accurate_2020,golze_core-level_2018,cavigliasso_accurate_1999,sen_accurate_2002,liu_benchmark_2019,lembinen_calculation_2020,tolbatov_comparative_2014,Hu_density_1996,zheng_performance_2019,pueyo_bellafont_performance_2016,pueyo_bellafont_performance_2016-1,besley_self-consistent-field_2009,klein_nuts_2021}. In addition, a few studies have considered the 2$p$ core levels of elements of the third row of the periodic table \cite{kahk_core_2021,ozaki_absolute_2017,kahk_accurate_2019,takahata_accurate_2010,aoki_accurate_2018,zhu_all-electron_2021,walter_offset-corrected_2016}.

Much less work has been carried out for systems containing transition metal atoms. A few studies applied the $\Delta$SCF approach to atomic systems: all subshells of neutral atoms with $Z$ between 2 and 106 were considered in \cite{huang_neutral-atom_1976}, and 2$s$ and 2$p$ binding energies in isolated transition metal ions were computed in \cite{clark_scf_1980}. Moreover, methods that go beyond the conventional $\Delta$SCF approach, where the wavefunction of the core hole state is expressed as a combination of multiple Slater determinants, have recently been used to study Ti 2$p$ photoemission from SrTiO3, and Fe 2$p$ photoemission from FeO and Fe$_2$O$_3$ by Bagus et al  \cite{bagus_new_2019,bagus_covalency_2020,bagus_combined_2021}. These studies are primarily concerned with explaining the satellite structures and peak broadenings observed in experimental measurements. Finally, we note that $\Delta$SCF calculations based on Hartree-Fock theory and complete active space SCF (CASSCF) calculations of Zn 2$s$ and 2$p$ core electron binding energies in the Zn atom, small molecular compounds of Zn, and small clusters of Zn metal and ZnO were reported by Rößler et al. in \cite{rosler_ab_2003,rossler_ab_2006}. 

In this work, we examine the performance of the all-electron $\Delta$SCF method based on density functional theory (DFT) for predicting 2$p_{3/2}$ binding energies in molecules containing first row transition metals. In particular, we carry out calculations for 60 different molecules containing Ti, V, Cr, Mn, Fe and Co using the formalism introduced in \cite{kahk_accurate_2019} and \cite{kahk_core_2021} and compare the calculated core electron binding energies to experimental values from gas phase photoelectron spectroscopy. We find that the discrepancy between calculated and measured binding energies is larger than for the lighter elements. Importantly, however, our results suggest that the error depends predominantly on the element and is rather insensitive to the atom's chemical environment. By correcting for this systematic error in the calculated binding energies, highly accurate absolute binding energies can be obtained.

\section{Methods}

All core electron binding energies reported in this work have been calculated using the $\Delta$SCF method, in which the binding energy is obtained as the total energy difference between the N-electron ground state and the N-1 electron state with a core hole. The total energy of the N-1 electron state is calculated by converging the self-consistent field in the presence of an explicit core hole, which is achieved by imposing a non-Aufbau-principle occupation of the Kohn-Sham eigenstates at each SCF iteration. The Maximum Overlap Method (MOM) is used to keep track of the core hole, in case the energy ordering of the core orbitals changes between successive SCF iterations \cite{gilbert_self-consistent_2008,michelitsch_efficient_2019}. The calculations have been performed using the formalism that was originally introduced in \cite{kahk_accurate_2019}. In particular, we use Density Functional Theory (DFT) with the Strongly Constrained, Appropriately Normed (SCAN) exchange-correlation functional \cite{sun_strongly_2015} and the spin-unrestricted open shell Kohn-Sham method. Scalar relativistic effects are accounted for via the scaled Zeroth Order Regular Approximation (scaled ZORA) \cite{chang_regular_1986,van_lenthe_relativistic_1994,faas_zora_1995}. To enable direct comparison with experimental measurements for the 2$p_{3/2}$ core electron binding energies, we subtract 1/3 of the experimental spin-orbit splitting from the calculated scalar-relativistic binding energy. We have used spin-orbit splittings of 6.4 eV, 7.7 eV, 9.7 eV, 11.2 eV, 13.1 eV, and 15.1 eV for the elements Ti, V, Cr, Mn, Fe, and Co respectively \cite{thompson_x-ray_2009}.

DFT calculations were carried out using the all-electron electronic structure program FHI-aims, in which the Kohn-Sham eigenstates are represented in terms of atom-centered basis functions defined on a numerical grid \cite{blum_ab_2009,havu_efficient_2009,knuth_all-electron_2015}. For the atoms with a core hole, the default "Tier" basis sets of FHI-aims were augmented with additional functions in order to facilitate the relaxation of all remaining core and valence electrons in the presence of a core hole. These core-augmented basis sets are provided in the Supplementary Information which also contains further details on basis set convergence.$^\dag$ For the remaining atoms, the default basis sets of FHI-aims at the "tight" level were employed \cite{blum_ab_2009}.  All $\Delta$SCF calculations were performed at the relaxed molecular geometries. For geometry relaxation, the SCAN functional, the "atomic ZORA" treatment of relativistic effects, and the default "tight" basis sets of FHI-aims were used \cite{blum_ab_2009}. The integration grids were tightened as follows: the radial multiplier was set to 8, and the outer angular grid was increased by one step.\cite{blum_ab_2009,delley_high_1996} Geometries were relaxed using the Broyden-Fletcher-Goldfarb-Shanno (BFGS) algorithm until the forces on all atoms were less than 5$\times$10$^{-3}$ eV \AA$^{-1}$ \cite{knuth_all-electron_2015,head_broydenfletchergoldfarbshanno_1985}. Sample input files are provided in the Supplementary Information.$^\dag$

\section{Results and Discussion}

\begin{figure}
\centering
  \includegraphics[width=8.8cm]{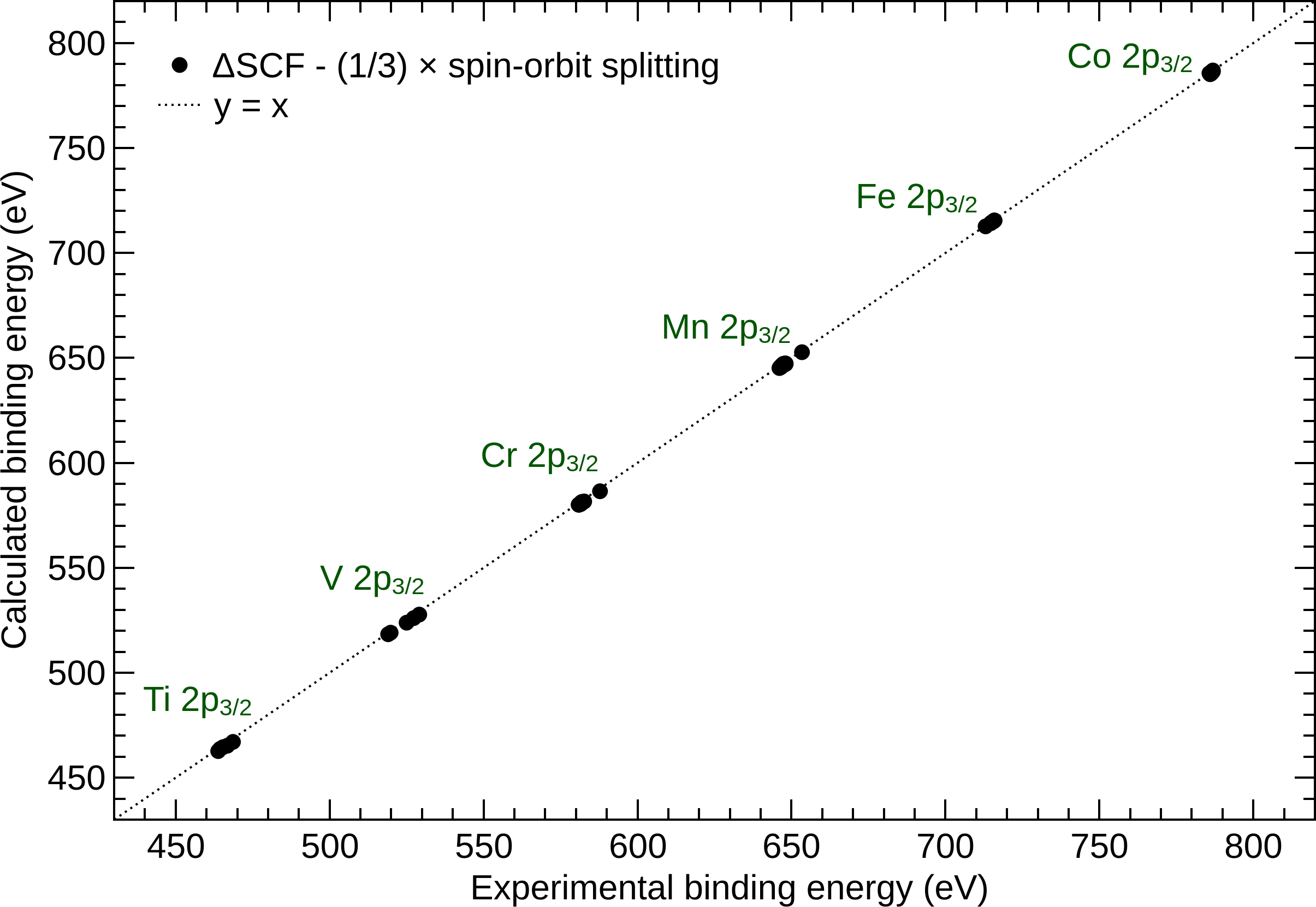}
  \caption{A comparison of the calculated core electron binding energies from scalar relativistic $\Delta$SCF calculations with spin-orbit correction to experimental values from gas phase photoelectron spectroscopy. The data for 60 molecular compounds containing 3d transition metals Ti, V, Cr, Mn, Fe and Co are shown together in one graph.}
  \label{Fig_elements_together}
\end{figure}

\begin{figure*}
\centering
  \includegraphics[width=17.4cm]{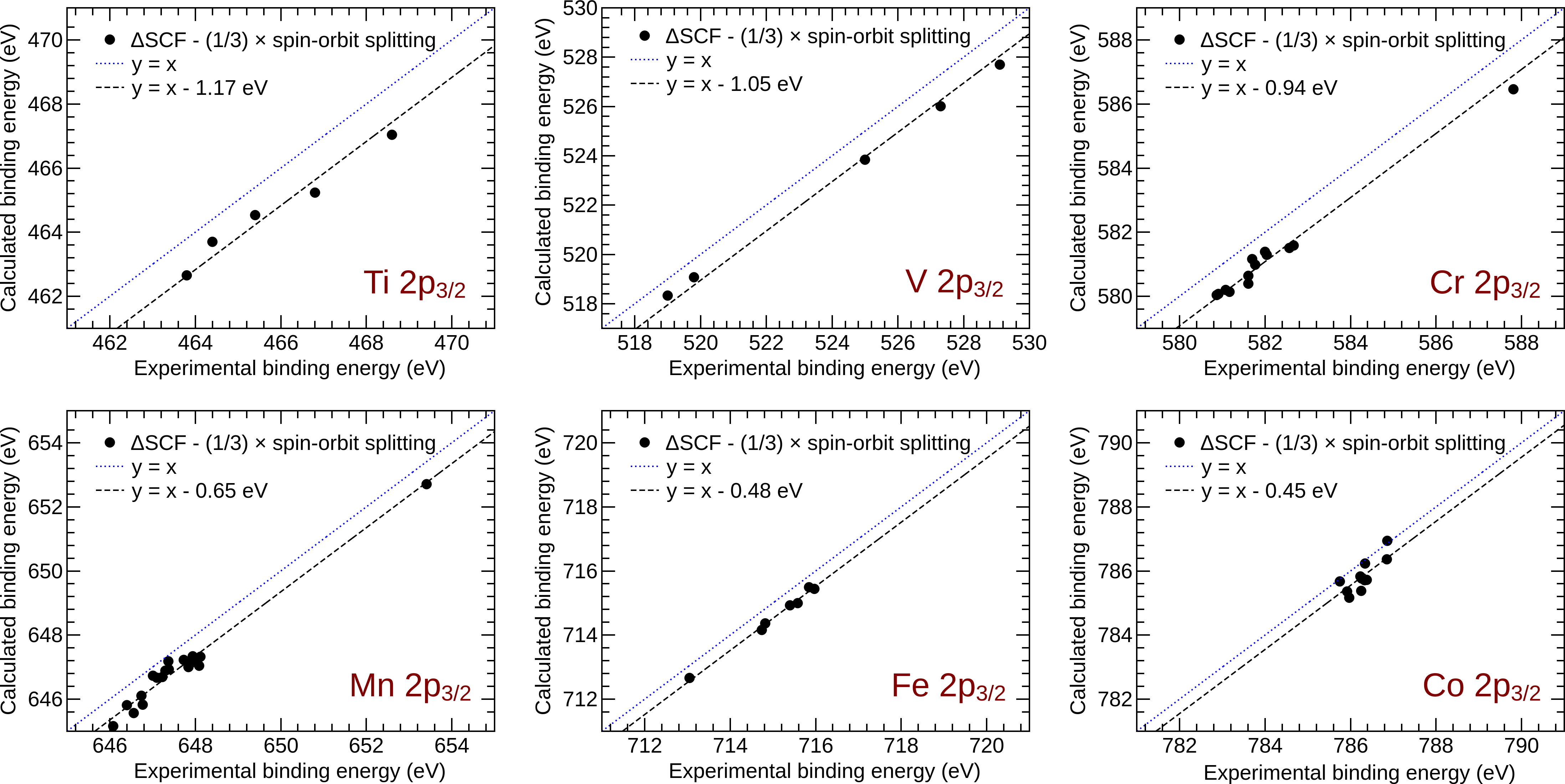}
  \caption{A comparison of calculated and experimental core electron binding energies. The blue dotted lines indicate perfect agreement between theory and experiment ($y=x$). The black dashed lines correspond to $y=x-a$, where $a$ is the mean error in the calculated core electron binding energies for that element.}
  \label{Fig_elements_separately}
\end{figure*}

A comparison of the calculated and experimental core electron binding energies for a dataset of 60 molecular transition metal compounds containing Ti, V, Cr, Mn, Fe, and Co is presented in Figures~\ref{Fig_elements_together} and \ref{Fig_elements_separately} and Table~\ref{Table_errors_for_whole_dataset}. Here, we only consider systems with a closed-shell ground state with most systems being organometallic compounds that follow the 18-electron rule. A full list of all molecules as well as all numerical results are provided in the Supplementary Information.$^\dag$ All experimental values have been taken from the compilation of gas phase X-ray Photoelectron Spectroscopy (XPS) data by Jolly and coworkers \cite{jolly_core-electron_1984}.

Figure~\ref{Fig_elements_together} shows the results for all molecules in one graph. It can be seen that the scalar relativistic all-electron $\Delta$SCF method with the spin-orbit corrections correctly captures the trend in 2$p_{3/2}$ binding energy with increasing atomic number across the series.

A closer look at the results for each individual element, see Figure~\ref{Fig_elements_separately}, indicates that the absolute values of the calculated core electron binding energies are always smaller than the measured value (with only one exception). The mean signed errors of the calculated core electron binding energies are -1.17 eV for Ti  2$p_{3/2}$, -1.05 eV for V  2$p_{3/2}$, -0.94 eV for Cr  2$p_{3/2}$, -0.65 eV for Mn  2$p_{3/2}$, -0.48 eV for Fe  2$p_{3/2}$, and -0.45 eV for Co  2$p_{3/2}$. These errors are significantly greater than the mean signed errors found for the 1s levels in the elements boron to fluorine, and the 2$p_{3/2}$ levels in the elements silicon to chlorine obtained in our previous study~\cite{kahk_accurate_2019}, which were all smaller than 0.35 eV, and typically below 0.15 eV. 

However, Fig.~\ref{Fig_elements_separately} also suggests that the errors of the absolute 2$p_{3/2}$ binding energies depend mostly on the element, but not significantly on the chemical environment of the transition metal atom. In particular, it can be seen that the results are well described by a shifted line $y=x-a$ (black dashed lines in Fig.~\ref{Fig_elements_separately}), where $y$ and $x$ denote the calculated and measured binding energies, respectively, and $a$ is the magnitude of the systematic error in the calculated core electron binding energies. Of course, the value of $a$ is equal to the mean signed error for each element (given in the previous paragraph). In other words, accurate absolute binding energies can be obtained once an element-specific correction is added to the calculated $\Delta$SCF binding energies. It should be noted that some of the corrections used in this work (e.g. for Ti and V) are currently derived from a relatively small number of datapoints. This is due to the limited availability of gas phase reference data for these elements. Future studies where a greater variety of compounds are studied, including solids and surface species, should be able to improve on this aspect.

\begin{figure*}
\centering
  \includegraphics[width=12cm]{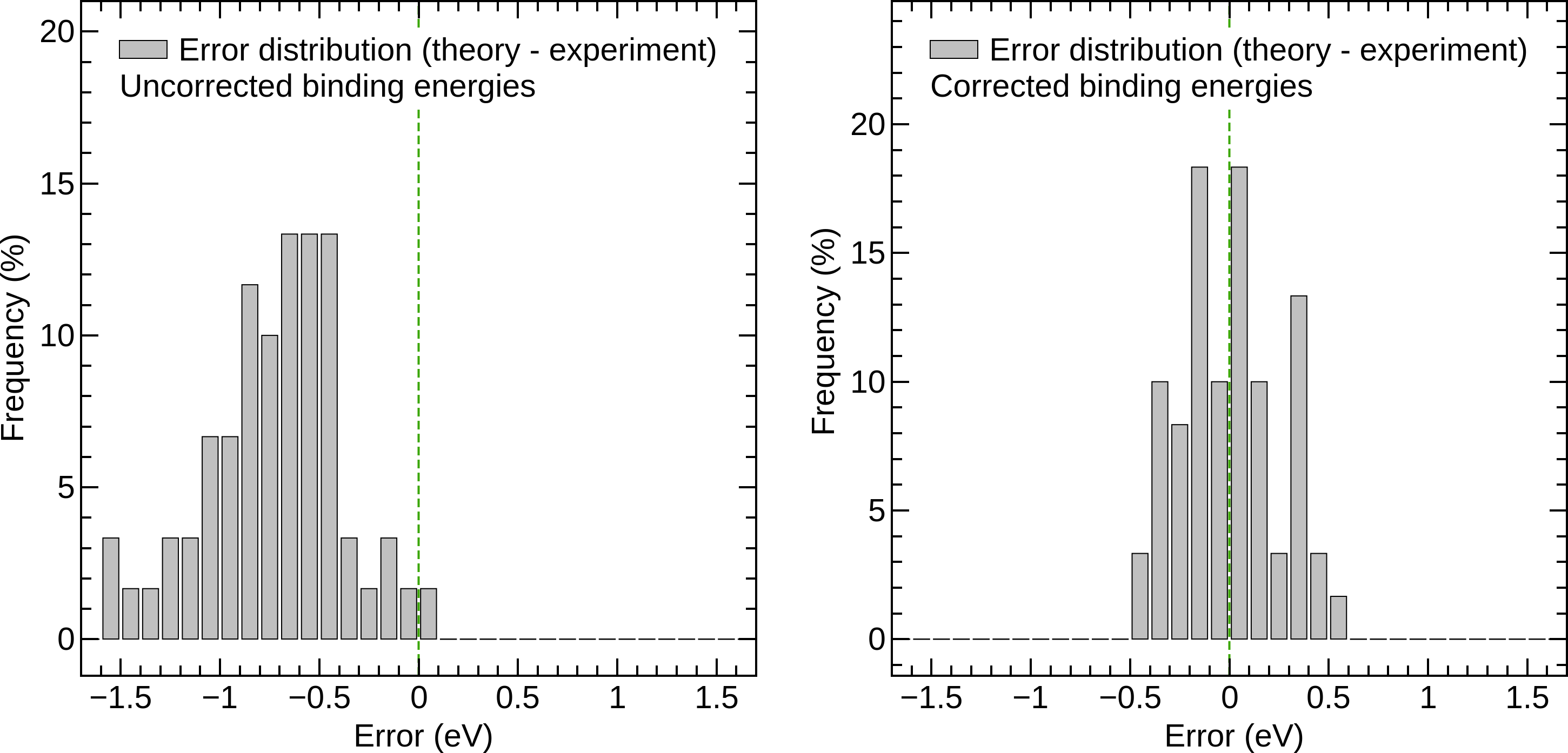}
  \caption{Histograms showing the distribution of errors in the calculated binding energies with and without the element-specific corrections.}
  \label{Fig_bar_graphs}
\end{figure*}

The accuracy of the calculated binding energies with and without the element-specific systematic corrections is analyzed in Table~\ref{Table_errors_for_whole_dataset} and Figure~\ref{Fig_bar_graphs}. Before the application of the corrections, the mean error for the whole dataset is -0.73 eV and the mean absolute error is 0.73 eV. In contrast, the mean absolute error in the binding energies after the application of the element-specific correction is just 0.20 eV. The distribution of errors of the corrected binding energies is shown in Fig.~\ref{Fig_bar_graphs}. In almost all cases, errors smaller than 0.5 eV are found.

\begin{table*}
\small
  \caption{The mean errors and mean absolute errors of the calculated 2$p_{3/2}$ binding energies with and without the element-specific correction. The mean errors of the corrected binding energies are not shown because they are zero by construction. All values are given in electronvolts.}
  \label{Table_errors_for_whole_dataset}
  \begin{tabular}{ l c c c c c c c}
    \hline
    & Ti 2$p_{3/2}$ & V 2$p_{3/2}$ & Cr 2$p_{3/2}$ & Mn 2$p_{3/2}$ & Fe 2$p_{3/2}$ & Co 2$p_{3/2}$ & Entire Dataset \\
    \hline
    Number of molecules & 5 & 5 & 13 & 19 & 7 & 11 & 60 \\
    Mean Error (uncorrected) & -1.17 & -1.05 & -0.92 & -0.65 & -0.48 & -0.45 & -0.73 \\
    Mean Absolute Error (uncorrected) & 1.17 & 1.05 & 0.92 & 0.65 & 0.48 & 0.47 & 0.73 \\
    Mean Absolute Error (corrected) & 0.32 & 0.28 & 0.19 & 0.20 & 0.07 & 0.24 & 0.20 \\
    \hline
  \end{tabular}
\end{table*}

In light of the excellent agreement of the calculated absolute core electron binding energies for the 2$p$ levels in lighter elements and the 1$s$ levels in the elements lithium to magnesium, one naturally wonders why significant element-specific corrections are needed to obtain accurate absolute binding energies for the  2$p_{3/2}$ core levels of first row transition metals. We have carefully considered the dependence of the calculated binding energies on the basis set used in the calculation, and found that basis set incompleteness cannot account for the observed systematic errors (see Supplementary Information for details).$^\dag$ Therefore, we can say that the two most important sources of error are the treatment of relativistic effects and the choice of the exchange-correlation functional.  

Our treatment of relativistic effects using scalar-relativistic DFT (followed by subtracting one third of the atomic spin-orbit splitting to obtain 2$p_{3/2}$ binding energies) is well justified for the 1s levels in elements of the second period of the periodic table and also for the 2$p$ levels of the elements silicon to chlorine even though the magnitude of the spin-orbit splittings are already significant in these systems (ranging from 0.4 eV for Si to 1.6 eV for Cl). In contrast, for the 2$p$ levels of the elements titanium to cobalt, the spin-orbit splittings are much greater, ranging from 6.4 eV for Ti to 15.1 eV for Co and therefore the applied spin-orbit correction shifts the binding energies by several eV. It is possible that for shifts of this magnitude, including the effects of spin-orbit coupling via a post-SCF correction is no longer reliable. In future studies, it would be interesting to examine whether fully relativistic $\Delta$SCF calculations, in which the energy of the final state with a  2$p_{3/2}$ core hole can be calculated directly, yield more accurate absolute core electron binding energies. Of course, the influence of the exchange-correlation functional on the accuracy of the calculated binding energies should also be investigated. 

\section{Conclusions}

Previous studies have established the $\Delta$SCF method as a highly accurate and computationally affordable method for calculating core-electron binding energies. However, thus far, the vast majority of theoretical work has focused on elements of the 2nd and 3rd periods of the periodic table, whereas experimentally, core level spectra of all stable elements, except for H and He, are studied. In this work, we have evaluated the suitability of the $\Delta$SCF method for predicting 2$p_{3/2}$ core electron binding energies in compounds of the first row transition metals Ti, V, Cr, Mn, Fe, and Co.

We have found that a formalism based on scalar-relativistic DFT calculations using the SCAN functional systematically underestimates absolute TM 2$p_{3/2}$ binding energies in these compounds, and that the errors in the calculated absolute core electron binding energies are significantly larger than the errors found for lighter elements using the same method \cite{kahk_accurate_2019}. However, we have also found, that for the dataset of 60 molecules considered in this work, the calculated relative binding energies are still highly accurate, with a mean absolute error of 0.20 eV compared to experiment. Moreover, we have shown that with the use of element-specific corrections, the prediction of accurate absolute TM 2$p_{3/2}$ core electron binding energies in closed shell compounds of first row transition metals is also possible.

Together with the recent work on extending the $\Delta$SCF method to periodic solids~\cite{kahk_core_2021}, these results lay the foundations for the use of $\Delta$SCF calculations for predicting core electron binding energies in a wide range of experimentally relevant systems. Further theoretical work is required to examine the possibility of performing constrained fully relativistic DFT calculations with the non-Aufbau principle occupation of eigenstates that would allow the total energies of core hole states, where the core hole resides in a spin-orbit split $p$-, $d$-, or $f$-orbital, to be calculated directly.

\section*{Conflicts of interest}
There are no conflicts to declare.

\section*{Acknowledgements}
This project has received funding from the European Union's Horizon 2020 research and innovation programme under grant agreement No 892943. JMK acknowledges support from the Estonian Centre of Excellence in Research project “Advanced materials and high-technology devices for sustainable energetics, sensorics and nanoelectronics” TK141 (2014-2020.4.01.15-0011). This work used the ARCHER2 UK National Supercomputing Service via J.L.’s membership of the HEC Materials Chemistry Consortium of UK, which is
funded by EPSRC (EP/L000202).

\end{document}